\documentclass{bioinfo}
\pdfoutput=1
\usepackage{hyperref}
\usepackage{amsmath}
\usepackage{graphicx}
\hypersetup{colorlinks=true, citecolor=blue, linkcolor=blue, urlcolor=black}

\begin{document}

\title[Pathway Discovery via Dynamic Sensitivity Analysis]{Intervention Pathway Discovery via Context-Dependent Dynamic Sensitivity Analysis}
\author[G.Zhou \textit{et~al}.]{Gaoxiang Zhou\,$^{\text{\sfb 1,}*}$, Kai-Wen Liang\,$^{\text{\sfb 2}}$ and Natasa Miskov-Zivanov\,$^{\text{\sfb 1}}$}
\address{$^{\text{\sf 1}}$Department of Electrical and Computer Engineering, University of Pittsburgh, Pittsburgh, 15260, United States and \\
$^{\text{\sf 2}}$Department of Electrical and Computer Engineering, Carnegie Mellon University, Pittsburgh, 15213, United States}
\corresp{$^\ast$To whom correspondence should be addressed. Email: \href{gaz11@pitt.edu}{gaz11@pitt.edu}}

\abstract{The sensitivity analysis of biological system models can significantly contribute to identifying and explaining influences of internal or external changes on model and its elements. We propose here a comprehensive framework to study sensitivity of intra-cellular networks and to identify key intervention pathways, by performing both static and dynamic sensitivity analysis. While the static sensitivity analysis focuses on the impact of network topology and update functions, the dynamic analysis accounts for context-dependent transient state distributions. To study sensitivity, we use discrete models, where each element is represented as a discrete variable and assigned an update rule, which is a function of element's known direct and indirect regulators. Our sensitivity analysis framework allows for assessing the effect of context on individual element sensitivity, as well as on element criticality in reaching preferred outcomes. The framework also enables discovery of most influential pathways in the model that are essential for satisfying important system properties, and thus, could be used for interventions. We discuss the role of nine different network attributes in identifying key elements and intervention pathways, and evaluate their performance using model checking method. Finally, we apply our methods on the model of naive T cell differentiation, and further demonstrate the importance of context-based sensitivity analysis in identifying most influential elements and pathways.}

\maketitle
\section{Introduction}
Understanding sensitivity is a critical step in studying system's adaptability to environment, and its robustness against perturbations, both of which are considered indispensable in living organisms (\citealp{1:shmu04}). In biological systems, the sensitivity assessment can indicate how susceptible the system (\textit{e.g.}, a cell) and its components (\textit{e.g.}, proteins or genes) are to internal or external changes. In turn, the analysis of sensitivity can help locate critical regulatory pathways in the system, and give insights into the robustness and responsiveness to interventions or treatments. To study the sensitivity in biological systems, researchers rely on carefully designed \textit{in vivo} or \textit{in vitro} experiments, or on analysis of computational models that can represent and mimic the behavior of the real systems \textit{in silico}. The sensitivity analysis approaches, and their applications, vary depending on studied systems and model types. In this work, we focus on the sensitivity of biochemical, intra- and inter-cellular networks, that is, signal transduction, metabolic, and gene regulatory networks (\citealp{1:shmu04}).

There are several different approaches to modeling biochemical networks, which depend on the size of the network, and even more, on the information available about network components and their interactions. A common approach uses ordinary differential equations (ODEs) derived from the reaction network (\citealp{2:faed09}). However, reaction networks grow exponentially with the number of network components (\textit{e.g.}, receptors, ligands, kinases, \textit{etc}.), and while fast methods are available to numerically solve large sets of ODEs, a common issue with this type of modeling is the lack of knowledge about network details necessary for accurate ODE-based models. It is often the case that we are familiar only with indirect cause-effect relationships for a number of interactions in the network, as there is no information (or existing knowledge) about exact mechanisms and the parameters necessary to create ODEs. Therefore, for such systems, a modeling approach is required which can accurately integrate all known interactions, and efficiently extend models with newly found interactions.

The discrete modeling approach has been suggested in the past to overcome the issue of missing information while still providing important insights into system behavior (\citealp{3:albe14}; \citealp{4:misk13}; \citealp{5:sun14}; \citealp{6:gan16}). Although based on relatively simple discrete formalism, this approach allows for automated integration of both prior knowledge and data into models. In this work, we focus on sensitivity in discrete models, in general, and logical or Boolean models, in particular.

In order to understand and explain the dynamics of biochemical network models, we use sensitivity analysis not only to study the influence from immediate, directly connected regulators, but also to explore global influences from indirectly connected elements. Previous research on sensitivity analysis in models of biochemical networks explored several directions. First, in (\citealp{1:shmu04}; \citealp{7:shmu02}; \citealp{8:shmu02}; \citealp{9:anon07}), the authors studied ``local'' average sensitivity of a model element and the influence of its regulators on this element, while (\citealp{10:qian09}; \citealp{11:trin16}) focused on the ``global'' long-run sensitivity of how likely a mutation is to change the converging attractor (steady-state) of a biological network. In this work, we start with the former direction to explore sensitivity ``locally'', and then extend the analysis by assigning a sensitivity value to each interaction in the network, and use these values to study sensitivity ``globally''. Second, considering in particular discrete models of biological networks, sensitivity analysis has been applied only on two types of models, the probabilistic Boolean networks (PBN), which are inferred from gene expression profiles (\citealp{7:shmu02}; \citealp{8:shmu02}; \citealp{10:qian09}) and random Boolean networks (RBN), randomized according to biological observations (\citealp{1:shmu04}; \citealp{11:trin16}; \citealp{12:koch14}). To the best of our knowledge, there has been no related sensitivity study on models assembled from individual element interactions using knowledge sources such as published literature, or information from experts, thus we refer to them as \textit{knowledge-based models (KBM)}. In contrast to PBN and RBN models, KBM models allow for various function types in network nodes (\textit{i.e.}, model elements), not only Boolean functions. Finally, previous research assumed static state distributions in studied models, such as uniform state distribution (\citealp{1:shmu04}; \citealp{9:anon07}; \citealp{12:koch14}) or steady-state distribution (\citealp{7:shmu02}; \citealp{8:shmu02}; \citealp{9:anon07}), neither of which captures transient context-dependent trajectories, and the dynamics of models between initial and steady state. In this work, we propose a more comprehensive sensitivity analysis framework, in which we define element influence and sensitivity with respect to the \textit{transient and scenario-dependent state distributions} of the system.

We perform both static and dynamic sensitivity analysis, the former assuming uniform state distribution, and the latter using a dynamic distribution estimated from stochastic simulation (\citealp{4:misk13}) trajectories under a particular scenario. Static sensitivity analysis can offer insights about the topological structure of the model and canalizing degree (\citealp{9:anon07}) of Boolean functions, but it entails no knowledge about the system beyond individual element update rules. Dynamic sensitivity analysis, on the other hand, incorporates the context-based bias into both individual and joint element state distributions at the intermediate points along trajectories. We propose a method to determine these joint distributions from \textit{dynamic simulation traces}, and we demonstrate the importance of performing dynamic analysis.

Our sensitivity analysis approach \textit{enables the extraction of globally important pathways}, which is critical in studying biological systems. Once we identify these pathways, we can design control strategies and interventions to tune system inputs or alter element values during the transient process. Through the construction of a weighted directed graph for the model, we have reduced the problem of extracting important paths to a minimum-cost graph search problem.

We also validate our sensitivity analysis results using the model checking approach (\citealp{13:misk16}; \citealp{14:wang16}) and demonstrate that element influence (especially in the case of dynamic analysis), outperforming many other network attributes, is a useful measure in pathway extraction. As part of the validation process, we propose a procedure to apply our sensitivity analysis framework to biological networks via the best-first search algorithm.

To demonstrate the advantages of our proposed work, we applied our sensitivity analysis framework on the previously published model of the intra-cellular network that controls naive T cell differentiation (\citealp{15:misk13}). With element influence and sensitivity measures, we evaluate, under different scenarios, individual elements and interactions, as well as the overall model. Finally, we highlight the significance of sensitivity analysis by comparing and contrasting observations from static network-based analysis and context-dependent dynamic analysis.

\section{Background and Related Work}
In this section, we provide the background of the modeling and simulation approaches that we use as part of our sensitivity analysis framework.

\subsection{Discrete Modeling Approaches}
\label{2.1:dma}
In discrete models, each element is assigned a set of discrete values, and an update rule, which is a function of its regulators. The model elements are, therefore, connected into a complex network via their update rules.

Among discrete modeling approaches, Bayesian Networks and Boolean Networks have been two common structures used in studying biological networks, in particular, gene regulatory networks (\citealp{16:kade08}; \citealp{17:miha01}; \citealp{18:peer01}; \citealp{19:davi08}). A Bayesian Network describes the probabilistic dependence interactions among elements. The model is assembled via structure learning (integrating prior knowledge from literature, experts) and parameter learning (\citealp{16:kade08}; \citealp{17:miha01}). However, it fails to capture cyclic interactions and neglects feedback loops (\citealp{20:frie00}). Boolean Networks (BN) (\citealp{16:kade08}), together with Probabilistic Boolean Networks (PBN) (\citealp{7:shmu02}; \citealp{8:shmu02}) and Random Boolean Network (RBN)s (\citealp{1:shmu04}; \citealp{11:trin16}), aim to model the biological phenomena with simple Boolean formalism, where each element is assigned a Boolean variable to represent its state, and one or multiple candidate Boolean update function. In the BN-based approaches, state transitions are usually assumed to be synchronous, that is, all elements are updated simultaneously. In contrast to Bayesian Networks, BNs allow modeling of cyclic regulation and intertwined feedback loops. On the other hand, BNs inferred from data are highly dependent on expression profiles and cannot incorporate prior knowledge of interactions between elements. In addition, specific to PBNs, their high computational complexity limits the application to large-scale networks.

The discrete modeling approach that we use in this work allows for feedback loops, integration of both prior knowledge and data, more realistic simulation of state transitions, as well as analysis of large hybrid networks that include protein-protein interactions, gene regulation, and even metabolic pathways and cellular processes.

\subsection{Modeling Preliminaries}
\label{2.2:mp}
The construction of a model begins with identifying key system components, and their interactions, usually through literature reading, data analysis or discussions with experts (\citealp{21:saye17}). The extraction of this information from knowledge sources allows modelers to identify the set of model elements $V=\{x_1,x_2,...,x_N\}$, where $N$ is the number of elements in the model. For each element $x_i$ in $V$, we define influence set $V_{IS}^i \subset V$ as the set of other elements that regulate it, as well as the polarity (positive or negative) of these regulations, and we also define the influence set size for element $x_i$ as $k(i)=|V_{IS}^i|$. The influence sets in a model can be illustrated as an influence graph $G(V,E)$, where nodes are model elements from $V$, and directed edges in $E$ are regulatory interactions between elements.

For a given element $x_i$, we can also define the number of discrete values representing different levels of the element's activity, $n_i$, such that $x_i \in X_i:\{0,1,2,...,n_i-1\}$. We will call the value assigned to element $x_i$ a \textit{state of element} $x_i$, and by assigning values to all elements in the model, we obtain overall \textit{model state}. Once we define model elements, their influence sets, and levels of activity, we create an \textit{executable model} that can be simulated, by assigning update functions to a subset (or all) of model elements. Update functions can be derived according to the information available about element regulations, as described in (\citealp{4:misk13}). Thus, an influence graph $G(V,E)$, together with additional information about regulatory functions, is used to create a discrete model, $M(V,F)$, with $F=\{f_1,f_2,...,f_N\}$, where, for each model element $x_i$, we define a deterministic discrete element update function, $f_i$, by mapping a $k(i)$-dimensional non-negative vector to a non-negative integer in the set $X_i$. Boolean models are considered a special case of discrete models where the domain of all elements is $B={0,1}$ and the primitive operators used to create any function in $F$ include AND, OR, and NOT. In Fig. \ref{fig:01}(a), we show an example three-element Boolean model, its influence graph, and element update functions.

\subsection{Model Simulation}
\label{2.3:ms}
\begin{figure}[thpb]
\centerline{\includegraphics[width=8cm]{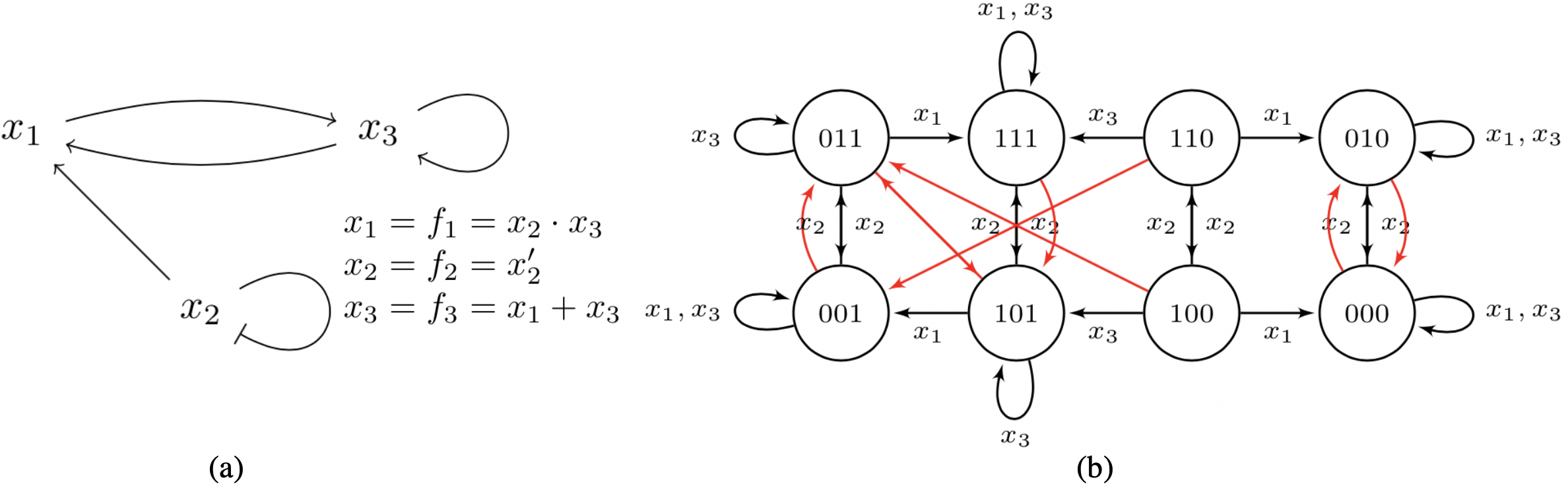}}
\caption{Discrete modeling approach and simulation: (a) A toy model $M(V,F)$ with model elements $V:\{x_1,x_2,x_3\}$ and their update functions; `` $\cdot$ '' denotes AND operator, where both regulators are required to be active (equal 1) to activate regulated element; ``+'' denotes OR operator, where either regulator being active is sufficient to activate regulated element; `` $'$ '' denotes NOT operator, where inactivation of regulator activates regulated element. (b) The state transition graph of the toy model under SMLN (red arrows) and RSB-SQ schemes (black arrows, labeled with updated element name).}\label{fig:01}
\end{figure}

Discrete models have been studied either formally (\citealp{13:misk16}; \citealp{22:wang16}) or using simulations (\citealp{23:saye17}). Since we focus in this work on discrete models $M(V,F)$, and also account for the randomness in the occurrence time of biochemical reactions, we use DiSH simulator (\citealp{23:saye17}). DiSH supports several simulation schemes, which can be divided into two groups: simultaneous (SMLN) and sequential (SQ). In the SMLN update scheme (commonly used for BN-based models), current state values of all elements are used to simultaneously compute their next state values, thus, the SMLN simulation scheme is deterministic. In the SQ update scheme, elements are updated sequentially, one after the other. The order of element updates is random, and the update rate for each element depends on the information available about its regulatory mechanisms. While DiSH simulator supports several SQ-based schemes, in this work we use the RSB-SQ (random step-based sequential) scheme, in which, at a given simulation step, only one randomly selected element is updated according to its update rule, while all other elements are not updated during that step. Fig. \ref{fig:01}(b) shows the state transition graph for our example model (Fig. \ref{fig:01}(a)) under both the SMLN and RSB-SQ update schemes.

In order to obtain a trajectory for the model, that is, state transitions of all elements between initial state and final state, we need to define a simulation \textit{scenario}. Each scenario includes information about initial values (states) of all non-input model elements (\textit{i.e.}, nodes in the influence graph that have arrows pointing at them), initial values for all model inputs, and, when available, perturbations that are assumed to happen at a particular model element, at a specified time point. The simulation is then executed following the scenario, from the initial state, until a pre-specified final state, which is indicated with the number of simulation \textit{steps}. One simulation \textit{run} provides a trajectory of each model element between an initial and a final state.

\section{Sensitivity Analysis}
\begin{figure}[thpb]
\centerline{\includegraphics[width=8cm]{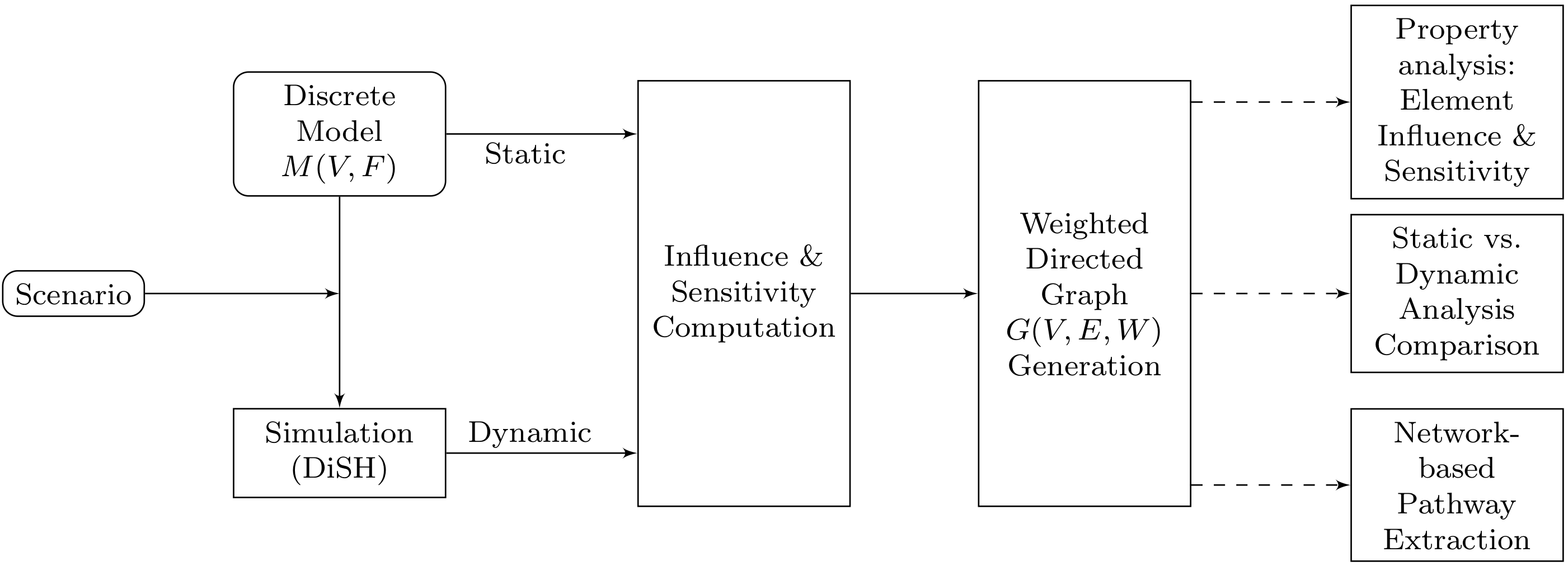}}
\caption{The flow diagram of our sensitivity analysis framework.}
\label{fig:02}
\end{figure}

In this section, we discuss the details of our methodology. We outline in Fig. \ref{fig:02} the data flow diagram of the framework. We use as inputs a model $M(V,F)$ and a scenario under which the model will be analyzed. The model alone is sufficient for static sensitivity analysis (described in detail in Section \ref{3.1:ei}, \ref{3.2:es}, \ref{3.3:sa}), while the scenario definition is required for dynamic sensitivity analysis (see Section \ref{3.4:da}). As mentioned in Section \ref{2.3:ms}, to obtain model trajectories for the dynamic sensitivity analysis, we run simulations using RSB-SQ scheme. We then define element influence and element sensitivity and show how to compute them. With the influence and sensitivity computation results, we can extend a discrete model graph $G(V,E)$ to a weighted directed graph $G(V,E,W)$, as described in Section \ref{3.5:saww}. An overall algorithm of our sensitivity analysis framework is also given in Section \ref{3.5:saww}.

\subsection{Element Influence}
\label{3.1:ei}
For a given set of model elements $V=\{x_1,x_2,...,x_N\}$, we define a vector $\boldsymbol{v}=(x_1,x_2,...,x_N)$ to represents the model's state. We are interested in computing a sensitivity of element $x_j$ to changes in the value of element $x_i$, where $i,j \in \{1,2,...,N\}$, and $i \neq j$. To find the sensitivity of function $x_j=f_j(x_1,x_2,...,x_N)$ to element $x_i$, that is, the influence of element $x_i$ on element $x_j$, we need to calculate the partial derivative of function $f_j$ with respect to element $x_i$. For logical models with Boolean variables and Boolean functions, the partial derivative is defined as (\citealp{1:shmu04}):
\begin{equation} \label{eqn1}
\frac{\partial f_j}{\partial x_i}=f_j|_{x_i=0} \oplus f_j|_{x_i=1}=f_{ji}(x_1,x_2,...,x_{i-1},x_{i+1},...,x_N)
\end{equation}
\noindent where $f_j|_{x_i=0}$ and $f_j|_{x_i=1}$ are the co-factors of $f_j$ with respect to $x_i$, which evaluate $f_j$ assuming $x_i=0$ and $x_i=1$, respectively.

It is obvious from Equation (\ref{eqn1}) that, $\frac{\partial f_j}{\partial x_i}$ does not depend on $x_i$, and only depends on other model elements $x_l \in V,l \neq i$, as determined from the update function $f_j$ of element $x_j$. Therefore, to find whether element $x_j$ can be influenced by $x_i$, we need to identify all possible values of vector $(x_1,x_2,...,x_{i-1},x_{i+1},...,x_N)$, for which partial derivative, $\frac{\partial f_j}{\partial x_i}$, is true (\textit{i.e.}, equal 1). It can be seen from Equation (\ref{eqn1}), that the partial derivative will be equal 1 if and only if, for given values of $(x_1,x_2,...,x_{i-1},x_{i+1},...,x_N)$, functions $f_j|_{x_i=0}$ and $f_j|_{x_i=1}$ have different values. In such cases, the value of function $f_j$ changes when $x_i$ changes, that is, element $x_i$ can influence element $x_j$.

The influence of element $x_i$ in function $f_j$ is defined as expectation of partial derivative of function $f_j$ with respect to element $x_i$:
\begin{equation} \label{eqn2}
\alpha_i^j=\alpha_i^{f_j}=E(\frac{\partial f_j}{\partial x_i})
\end{equation}
\noindent Since the partial derivative $\frac{\partial f_j}{\partial x_i}$ itself is a Boolean function with two possible values \{0,1\}, we can consider the expectation $E(\frac{\partial f_j}{\partial x_i})$ as a probability $P\{(x_1,x_2,...,x_{i-1},x_{i+1},...,x_N)|\frac{\partial f_j}{\partial x_i}=1\}$, that is, a ratio of the number of vectors $(x_1,x_2,...,x_{i-1},x_{i+1},...,x_N)$ for which $\frac{\partial f_j}{\partial x_i}=1$ to the number of all possible vectors $(x_1,x_2,...,x_{i-1},x_{i+1},...,x_N)$. This quantitatively describes the relationship between the regulator $x_i$ and regulated element $x_j$, as it provides the probability that a change in element $x_i$ will change the value of $x_j$.

In general, the procedure of calculating the influence of variable $x_i$ on variable $x_j$ (\textsc{ELEMENT\_INFLUENCE}($x_i,x_j$) in the algorithm described in Section \ref{3.5:saww}), can be divided into two main steps:

\textbf{(1)} \textsc{COMPUTE\_INFLUENTIAL\_VECTORS}($f_j,x_i$), which takes as arguments update function $f_j$ and one of its regulator variables $x_i$, returns the set of vectors for which the change in $x_i$ leads to a change in $f_j$;

\textbf{(2)} \textsc{EST\_PROB}($\boldsymbol{v}$), which computes the probability of occurrence of a vector of element values $\boldsymbol{v}$ from simulation trajectories; as default probability, we assume that values follow uniform distribution.

\begin{figure}[thpb]
\centerline{\includegraphics[width=8cm]{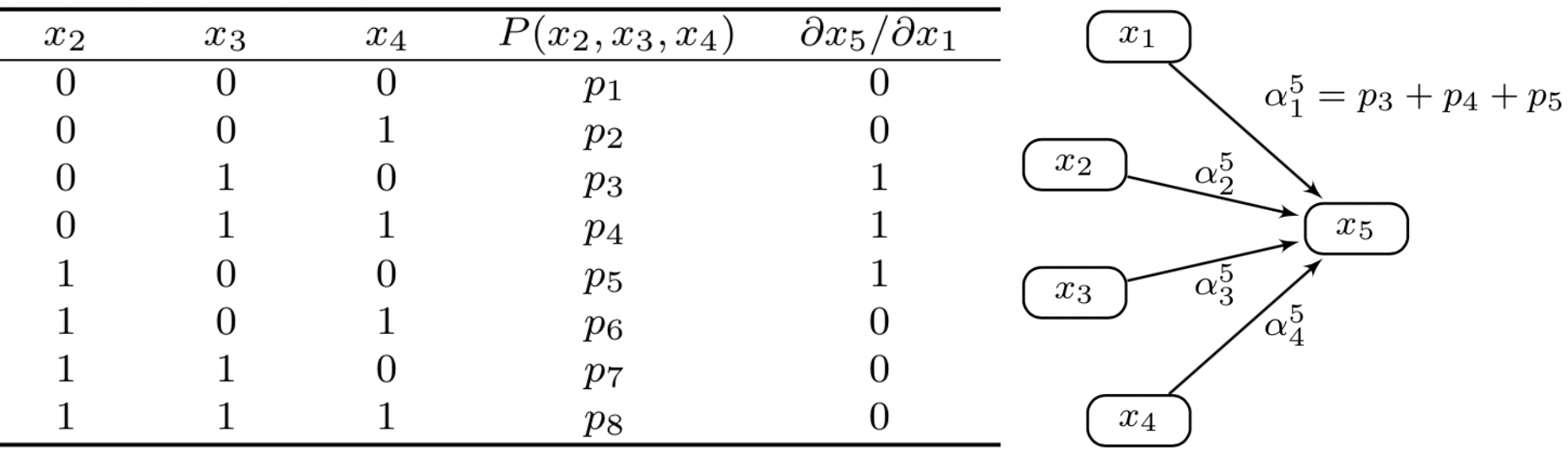}}
\caption{Element influence calculation of function $x_5=f_5=x_1x_2+x_1^{'}x_3+x_2 x_3^{'}x_4$.}
\label{fig:03}
\end{figure}

\textit{Example 1}: Given a Boolean function $x_5=f_5=x_1x_2+x_1^{'}x_3+x_2x_3^{'}x_4$, how can we determine the influence of element $x_1$ on $x_5$? First, we find all combinations of $x_2,x_3,x_4$ for which $\frac{\partial f_5}{\partial x_1}=x_2x_3^{'}x_4^{'}+x_2^{'}x_3=1$. As can be seen from the truth table in Fig. \ref{fig:03}(left), these are the value combinations that have probabilities of occurrence $p_3, p_4$, and $p_5$. Next, we add the probabilities of these input vectors and compute the influence $\alpha_1^5=p_3+p_4+p_5$. Similarly, we compute all other influences, $\alpha_2^5,\alpha_3^5,\alpha_4^5$ (listed in Fig. \ref{fig:03}(right)).

\subsection{Element Sensitivity}
\label{3.2:es}
Besides evaluating influence of individual regulators, we also investigate overall sensitivity of elements with respect to the combined influence of all their regulators (\textit{i.e.}, their influence sets). For a given element $x_j$, and its influence set $\{x_{j_1},x_{j_2},...,x_{j_{k(j)}}\}$, where $k(j)$ is the size of the influence set of $x_j$, we can decrease the size of vector $\boldsymbol{v}$ to only include $x_j$'s influence set elements, such that $\boldsymbol{v}=(x_{j_1}, x_{j_2},...,x_{j_{k(j)}})$. Obviously, there are $k(j)$ ways to flip only one value in vector $\boldsymbol{v}$. Each value flip could potentially change the value of $f_j$. In some cases, $f_j$ can be very sensitive to its input vectors (\textit{e.g.}, the sensitivity of function $f=x_{j_1} \cdot x_{j_2} \cdot...\cdot x_{j_{k(j)}}$ to input vector (1,1,...,1)). To take this into consideration, for a given element $x_j$, we define its element sensitivity to a certain value $\mathbf{v}$ of its influence vector $\boldsymbol{v}, s^j(\boldsymbol{v}=\mathbf{v})$ as the number of different ways individual elements of the influence set can be flipped within $\boldsymbol{v}$ to change the value of $x_j$. If we define $\boldsymbol{v}^i$ as a vector of size $k(j)-1$, same as vector $\boldsymbol{v}$ but missing element $x_i$, then we can compute $s^j(\boldsymbol{v}=\mathbf{v})$ as:
\begin{equation} \label{eqn3}
s^j(\boldsymbol{v}=\mathbf{v})=\sum_{i=1}^{k(j)} \frac{\partial f_j}{\partial x_i} (\boldsymbol{v}^i=\mathbf{v}^i)
\end{equation}

We then define an overall element sensitivity of $x_j$, to all possible influence set value vectors, with respect to a given value distribution, as:
\begin{equation} \label{eqn4}
s^j=E(s^j(\boldsymbol{v}=\mathbf{v}))=\sum_{i=1}^{k(j)} E(\frac{\partial f_j}{\partial x_i} (\boldsymbol{v}^i=\mathbf{v}^i)) = \sum_{i=1}^{k(j)} \alpha_i^j
\end{equation}

As shown in Equation (\ref{eqn4}), $s^j$ has been proven equivalent to the sum of the influences of all its regulators in function $f_j$ (\citealp{1:shmu04}), where individual regulator influences can be computed according to Equation (\ref{eqn2}). We later call this procedure of computing element sensitivity for $x$ as \textsc{ELEMENT\_SENSITIVITY}($x$).

We could see that the upper bound of $s^j$ is $k(j)$ as element influence $\alpha_i^j$ is bound by 1, and that the sensitivity of element $x_j$ is also dependent on the size of its influence set $k(j)$. For the same network as in \textit{Example 1}, the element sensitivity of $x_5$ to input vector $(x_1,x_2,x_3,x_4)$ is given by $\alpha_1^5+\alpha_2^5+\alpha_3^5+\alpha_4^5$.

As shown in (\citealp{1:shmu04}) and (\citealp{9:anon07}), if an element's sensitivity is greater than 1, this would lead to instability of this element (\textit{i.e.}, it enables this element's change to propagate out of control). Additionally, (\citealp{1:shmu04}) has shown that, even with the same element sensitivity, unbalanced influence distribution of its regulators can make elements behave more stable and robust than elements with balanced regulator influence distribution.

\subsection{Static Analysis}
\label{3.3:sa}
In this section, we describe our approach to analyzing static sensitivity of model elements, with respect to the distribution of all possible model states. If we assume a uniform distribution of model states, the influence of element $x_i$ in the regulation of $x_j$ can be expressed as
\begin{equation} \label{eqn5}
\alpha_i^j=\alpha_i^{f_j}=E(\frac{\partial f_j}{\partial x_i})=\frac{1}{2^{k(j)}}\sum_{\boldsymbol{v}} \frac{\partial f_j(\boldsymbol{v})}{\partial x_i}
\end{equation}

In general, static analysis provides a method to estimate the distribution of $\boldsymbol{v}$. In our algorithm shown in Section \ref{3.5:saww}, we denote this calculation as \textsc{EST\_PROB}($\boldsymbol{v}, static$), where $\boldsymbol{v}$ is any vector and $static$ is an indicator of the method. As can be seen from Equation (\ref{eqn4}), the sensitivity of model elements to changes in values of other elements is determined by the fixed set of element update rules. In other words, the static sensitivity analysis approach relies solely on $M(V,F)$, therefore, it provides information only about the connectivity and logic rules of the model. Thus, for the model in \textit{Example 1}, if we assume that $p_1=p_2=...=p_8=1/8$, $\alpha_1^5$ will be equal 3/8, which is solely determined by $\frac{\partial f_5}{\partial x_1}=x_2x_3^{'}x_4^{'}+x_2^{'}x_3=1$. Under the same assumption, we can easily compute the other three influences for \textit{Example 1} as $\alpha_2^5=5/8, \alpha_3^5=3/8$, and $\alpha_4^5=1/8$.

However, the assumption that the states are uniformly distributed fails to capture the dynamics of changes in model states, and consequently, the details about transient behavior along trajectories between the initial and final model states.

\subsection{Dynamic Analysis}
\label{3.4:da}
In a real biological system, many combinations of element states may never occur under particular conditions, or, given the structure of the system, they are not even possible. Additionally, scientists are often focusing on the system's transient response to interventions (\textit{e.g.}, drugs or treatments), since early response states could determine long-term outcomes (\citealp{15:misk13}). Therefore, in conducting sensitivity analysis, we account for (i) uncertainty in the information available about the system, (ii) influence of the system's initial state on its response, and (iii) stochasticity in system's response to stimulations and interventions. In other words, we assume that (1) the available information may not be sufficient to derive the exact state distribution, and that (2) the distribution of states may vary, depending on initial states, interventions and perturbations.

To tackle the challenge (1) above, we estimate the distribution of states through simulations. We can simulate the model for a pre-determined number of steps, and we use element trajectories that we obtain through simulations to derive the state preference (\textit{i.e.}, which states are more likely to be reached). To tackle the challenge (2) above, we conduct simulations for all the initial states that are of interest when studying a given system.

As described in Section \ref{2.3:ms}, DiSH simulator supports several different simulation schemes for discrete models. Here, we are interested in the random order update scheme (RSB-SQ), where simulation is run multiple times, on the same model and under the same scenario, and the trajectories obtained vary between different runs. Additionally, due to the randomness of trajectories, a particular state can be reached multiple times within a single run. Thus, we define the \textit{preference}, $pe(\boldsymbol{v})$, of a certain state as a ratio of the number of trajectories $n(T_{\boldsymbol{v}})$ that reached this state at least once, to the total number of trajectories $n(T)$. Finally, we estimate the probability $p(\boldsymbol{v})$ of a certain state $\boldsymbol{v}$ via normalization of $pe(\boldsymbol{v})$, that is:
\begin{equation} \label{eqn6}
pe(\boldsymbol{v})=\frac{n(T_{\boldsymbol{v}})}{n(T)}
\end{equation}
\begin{equation} \label{eqn7}
p(\boldsymbol{v})=\frac{pe(\boldsymbol{v})}{\sum_{\boldsymbol{v}} pe(\boldsymbol{v})}
\end{equation}

The dynamic trajectories obtained from simulation are often highly dependent on the initial state, and thus, the state preference results obtained from the sample trajectories, as well as our final estimated state distribution will be different for different initial states. In the algorithm (Fig. \ref{fig:05}), we name the procedure of estimating state probabilities in the dynamic-based analysis as \textsc{EST\_PROB}($\boldsymbol{v}$, $dynamic$, \textsc{TRAJECT}). This procedure takes as arguments an influence vector $\boldsymbol{v}=(x_{j_1}, x_{j_2},...,x_{j_{k(j)}})$, indicator variable $dynamic$, and a trajectory \textsc{TRAJECT}, obtained from simulations. Furthermore, this procedure could be divided into two steps, \textsc{COUNT}($\boldsymbol{v}$, \textsc{TRAJECT}) which returns the count of trajectories that reach state $\boldsymbol{v}$ (at any simulation step, not only the final step), and \textsc{NORMALIZE}(\textsc{COUNT}), which normalizes the counts according to the trajectory set size and scales them to sum to 1.

In the dynamic sensitivity analysis approach, the influence of element $x_i$ in function $f_j$ is defined by considering the state distribution $p(\boldsymbol{v})$.
\begin{equation} \label{eqn8}
\alpha_i^j=\alpha_i^{f_j}=E(\frac{\partial f_j}{\partial x_i})=\sum_{\boldsymbol{v}} \frac{\partial f_j(\boldsymbol{v})}{\partial x_i}p(\boldsymbol{v})
\end{equation}

For the model in \textit{Example 1}, we denote the influence vector of $x_5$ as $\boldsymbol{v}=(x_1,x_2,x_3,x_4)$. For instance, the initial state of the network we are interested in is $(x_1,x_2,x_3,x_4 )=(0,0,0,0)$, and RSB-SQ scheme will be applied. We run $n(T)=1500$ times to obtain 1500 trajectories, of which $n(T_{\boldsymbol{v}})=1050$ trajectories have reached state $(x_1,x_2,x_3,x_4)=(0,0,1,1)$. Then $pe(0,0,1,1)=1050/1500=0.7$. If, for example, we compute $\sum_{\boldsymbol{v}} pe(\boldsymbol{v})=7$, $p(0,0,1,1)$ could be obtained via normalization as $0.7/7=1/10$, no longer $1/8$ as static analysis. Element influences $\alpha_1^5,\alpha_2^5,\alpha_3^5,\alpha_4^5$ will then be calculated based on these new joint probabilities.

As can be seen from Equation (\ref{eqn8}), $\alpha_i^j$ depends on both state dynamics along the trajectories and the topological structure of the model. Given these dependencies, it is possible that some influence vector values that have significant influence under static analysis, never occur under dynamic analysis, thus leading to $\alpha_i^j=0$ in the dynamic case.

\subsection{Sensitivity Analysis with Weights}
\label{3.5:saww}
\begin{figure}[thpb]
\centerline{\includegraphics[width=7.7cm]{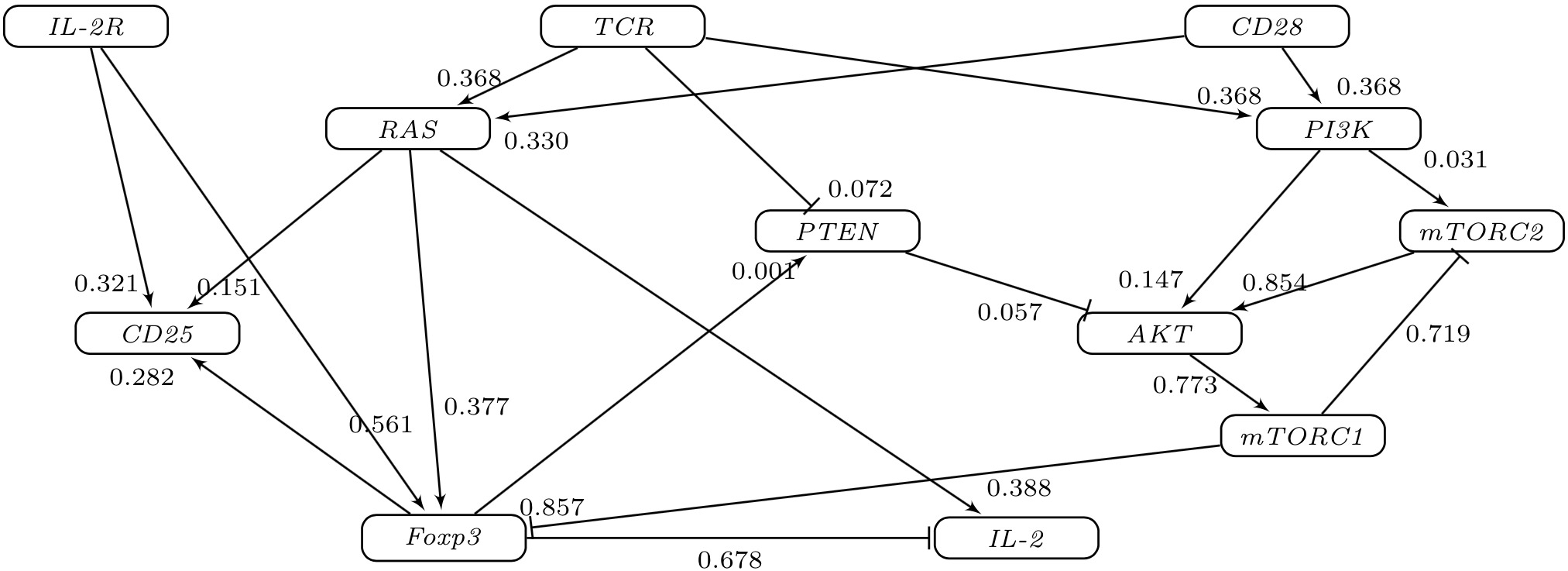}}
\caption{A weighted directed graph of a biological model (\citealp{15:misk13}) under certain scenario.}
\label{fig:04}
\end{figure}

With our sensitivity analysis framework, we are now able to extend the model influence graph $G(V,E)$ to a weighted directed graph $G(V,E,W)$ by adding weights $w_{ij}=\alpha_i^j$ to the directed edges pointing to $x_j$ from $x_i$. Here $\alpha_i^j$ can be obtained either from static analysis, as shown in Equation (\ref{eqn5}) or from dynamic analysis, as shown in Equation (\ref{eqn8}). In Section \ref{4.2:na}, we will also discuss several different graph attributes, and define weights based on those attributes. While for a pre-defined discrete model, the interaction map $G(V,E)$ is fixed, the weighted directed graph $G(V,E,W)$ can vary with respect to $p(\boldsymbol{v})$ of state $\boldsymbol{v}$, which is determined by a simulation scenario. We denote the generation of this graph as \textsc{GENERATION\_WDG}($M$, \textsc{ELEMENT\_INFLUENCE}($x_i,x_j$)), where $M$ is the model and \textsc{ELEMENT\_INFLUENCE}($x_i,x_j$)) provides the weights $w_{ij}=\alpha_i^j$. To illustrate this, we show in Fig. \ref{fig:04} an example graph $G(V,E,W)$ of the model from (\citealp{15:misk13}), using RSB-SQ simulation scheme and high antigen dose scenario.

\begin{figure}[thpb]
\centerline{\includegraphics[width=6.7cm]{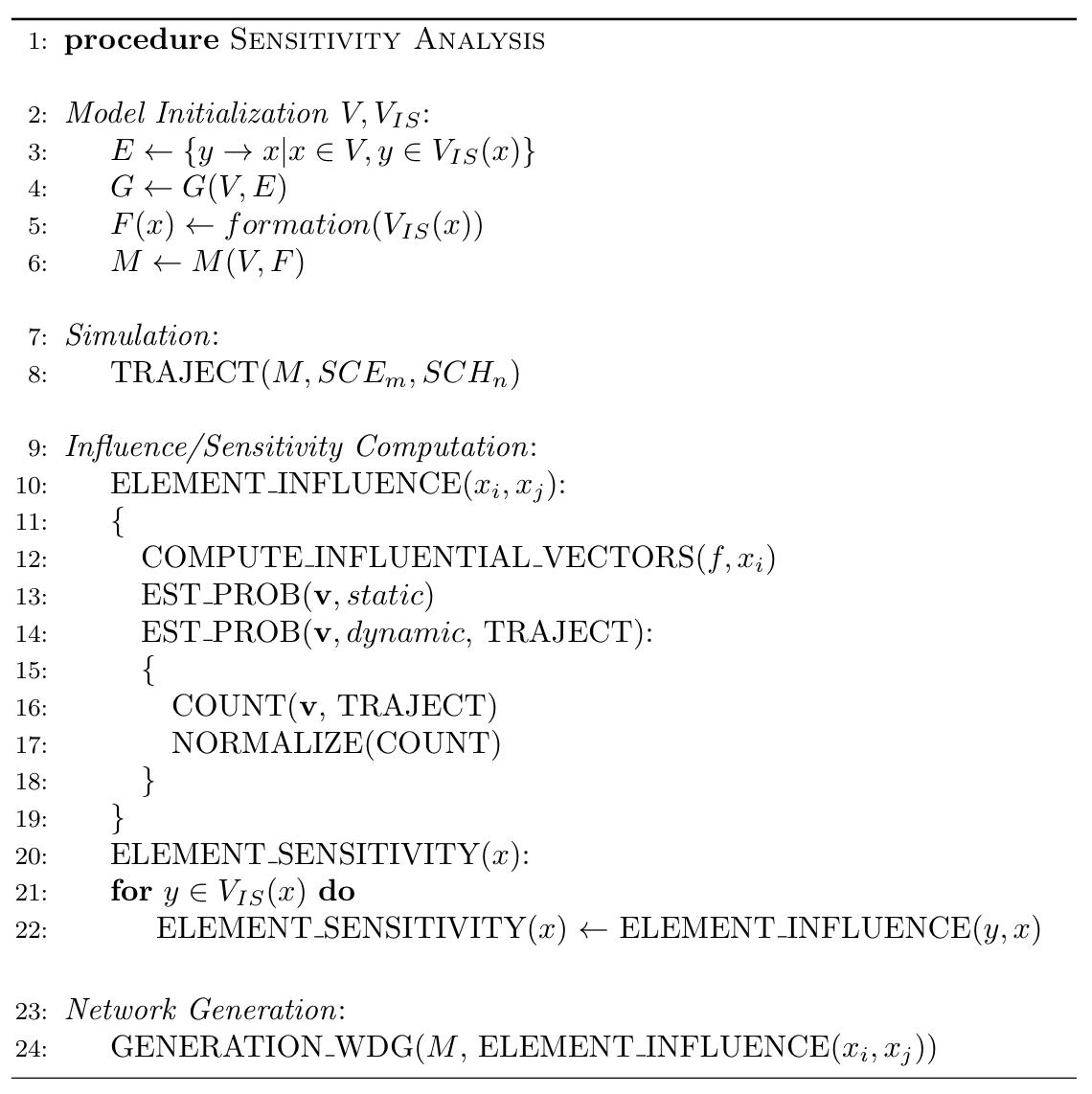}}
\caption{An overall algorithm of our sensitivity analysis framework}
\label{fig:05}
\end{figure}

In Fig. \ref{fig:05}, we provide the algorithm of our sensitivity analysis method, with standardized procedure of generating a weighted directed graph for a given model with Boolean update functions.

There are four main parts of the algorithm outlined in Fig. \ref{fig:05}. Lines 2-6 include model initialization, which returns an influence graph $G(V,E)$ and an executable model $M(V,F)$. Lines 7-8 represent the simulation part: given any scenario of interest, $SCE_m$, and any available scheme, $SCH_n$, it returns a trajectory set, \textsc{TRAJECT}. Lines 9-22 serve as the main part of the algorithm to compute influence and sensitivity. Lines 23-24 generate a weighted directed graph.

\section{Intervention Pathway Discovery}
In this section, we discuss application and validation of our sensitivity analysis approach in the discovery of intervention pathways using graph search. Specifically, we propose a method for finding intervention pathways, while emphasizing the importance of element influence and sensitivity attributes in the discovery process.

When studying biological systems, finding the most influential pathways from one component (source) of a system to another (target) is critical in understanding the system, and in developing effective interventions or treatments. Once we identify these pathways in a model, we can easily control and guide the model by tuning its inputs, or by deciding which elements we should change during the transient process. In Section \ref{4.1:bfs}, we first reduce the problem of extracting pathways to a minimum cost search problem, based on a given definition of importance. Here, we selected to use the best-first search (\citealp{24:heur84}) since it cooperates well with weighted graph and avoids unnecessary expansion via priority queue and hash table. In Section \ref{4.2:na}, we list nine candidate attributes that we will use in studying the model graph, and we define the weights that correspond to these attributes. To validate our proposed approach, in Section \ref{4.4:pee}, we modify a well-studied biological model (\citealp{15:misk13}) by removing the discovered important pathways, and then test the modified model to determine whether the model properties have been negatively affected. By testing whether the reduced model satisfies ground truths, we show that model properties are not preserved when we remove the pathways that our algorithm found.

\subsection{Best-First Search}
\label{4.1:bfs}
Given a weighted directed graph, a source node and a target node, our goal is to find the most ``important'' path(s) from source to target. The importance of a path could be interpreted in different ways. First, we could assume that the path having the highest cumulative element sensitivity is the most important, since the changes of regulators on this pathway will usually result in significant changes of regulated elements. Second, elements that have large influence sets usually serve as hubs in the network, and often form cycles (\citealp{25:albe05}), therefore, we could also define the most important pathway as the one passing through elements with large influence sets. In other words, given a particular definition of pathway importance, the extraction of important pathways could be reduced to problems of maximizing/minimizing some index along pathways in a graph search.

That is, given graph $G(V,E,W)$, source node $x_s$ and target node $x_t$, we aim to find a path that has the minimum weight summation (also known as cost) along it, that is, we define \textit{optimal path} as:

\begin{equation} \label{eqn9}
\{x_s,x_{p_1},...,x_{p_m},x_t\}\triangleq \operatorname*{arg\,min}_{x_s,...,x_t} w_{sp_1}+\sum_{i=1}^{m-1} w_{p_ip_{i+1}}+w_{p_mt}
\end{equation}

\noindent where $w_{sp_1}$ is the weight on the edge between $x_s$ and $x_{p_1}$, $w_{p_mt}$ is the weight on the edge between $x_{p_m}$ and $x_t$, and $w_{p_ip_{i+1}}$ is the weight associated with the edge pointing to node with index $p_{i+1}$ from node with index $p_i$. In the following section, we will discuss several approaches for computing $w_{ij}$. For instance, the definition in Section \ref{3.5:saww} is one of our definitions of weight.

We apply the best-first search via maintaining a tree of paths originating at the source node and extending those paths one edge at a time until the target is reached. At each node, the search algorithm needs to determine which of its paths to extend based on the cost of the partial path. Specifically, best-first search selects the path that minimizes $g(x)$, where $x$ is the next node on the path, and $g(x)$ is defined as the cost of the path from the source node to node $x$. Our algorithm of best-first search uses a priority queue to perform the repeated selection of nodes with minimum $g$ value. This priority queue is known as the open set. At each step, the node with the lowest $g$ value is removed from the queue, the $g$ values of its successors are updated accordingly, and these successors are then added to the queue. The algorithm continues until the target node is popped up from the queue. The trace of elements along which we reach the target is the path with minimum cost, which is also the most important path based on our interpretation of importance.

We also introduce the following in our best-first search algorithm: (i) the flexibility to allow or prohibit cycles as paths with cycles within a biological system form feedback or feed-forward controls; (ii) the flexibility to return one or multiple minimum cost paths as two paths might have the same minimum cost, and in addition, multiple paths are useful for later evaluation which we will elaborate in Section \ref{4.4:pee}.

\subsection{Network Attributes}
\label{4.2:na}
Different interpretations of importance will assign different weights to edges. The influence graph $G(V,E)$ and the discrete model $M(V,F)$ together carry a number of structural and dynamical features that can serve as useful attributes for detecting important pathways. We propose nine different attributes and define the edge weight according to the minimum cost graph search.

\textsc{IN-DEGREE}: $a_1(x)$, is the size of the $x$'s influence set. In the context of logical functions (and logic circuits), an element with large influence set tends to have a larger fan-in (the part of the circuit feeding into it), and dominates the circuit behavior. Accordingly, we aim to obtain path with more nodes with high in-degree (equivalently, low value in the reciprocal value of in-degree). Under such interpretation, we define the weight from node $x_i$ to node $x_j$ as $w_{ij}=\frac{1}{a_1(x_j)}$.

\textsc{OUT-DEGREE}: $a_2(x)$, counts how many times an element $x$ occurs in other elements' influence sets. Note that a self-loop regulation also counts. Generally, elements with higher out-degree also have a larger fan-out (the part of the circuit that this elements feeds into), and propagate the influence more broadly through the network. Thus, such elements often play hub roles in signal transduction. Similarly, we define $w_{ij}=\frac{1}{a_2(x_j)}$.

\textsc{SHORTEST\_LINK}($x$): $a_3(x)$, is defined as the shortest length path that goes through $x$ and connects given source and target elements. It is obvious that an element $x$ showing in a short path linking the source and target elements has a more significant impact in the relationship between source and target. Therefore, we could define $w_{ij}=a_3(x_j)$.

\textsc{LOOP\_COUNT}($x$): $a_4(x)$, counts how many loops go through $x$. In biology, feedback loops are common and often have crucial influence on behavior. A minor change in any element within a feedback loop could be amplified by multiple cycles through the loop. Thus, we expect that paths with elements that belong to multiple loops will be more influential, and therefore, we compute the weight as $w_{ij}=\frac{1}{a_4(x_j)}$.

\textsc{NON-BIAS}($x$): $a_5(x)$, defined as $1-2\cdot|Pr\{x=1\}-0.5|$ for a Boolean variable node. It is obvious that $a_5(x)$ ranges from 0 to 1. If the state of an element is biased towards 0 or 1 (\textit{i.e.}, non-bias degree approaches 0), this element is robust against perturbations and in most cases, it prevents further signal propagation (\citealp{1:shmu04}, \citealp{9:anon07}). We define $w_{ij}=-log(a_5(x_j))$ with the purpose of accumulating non-bias degree along the path.

\textsc{ELEMENT\_INFLUENCE}$(x_i,x_j)$: $a_6(x_i,x_j)$, a result of Equations (5) and (8), this attribute is further categorized as $a_{6-static}$ and $a_{6-dynamic}$, which return the static/dynamic analysis of influence of $x_i$ in the regulation of $x_j$, respectively. We define  $w_{ij}=-log(a_6(x_i,x_j))$.

\textsc{ELEMENT\_SENSITIVITY}($x$): $a_7(x)$, a result of Equation (\ref{eqn4}). Similar to $a_6$, this attribute could also be categorized into $a_{7-static}$ and $a_{7-dynamic}$, which return the static and dynamic analysis of element sensitivity of $x$, respectively. We define $w_{ij}=-log(a_7(x_j))$.

The attributes listed above could be divided into two types: structure-related and dynamics-related. The former category highlights the role of an element in network topological structure. Attributes $a_1,a_2,a_3,a_4,a_{6-static},a_{7-static}$ fall into this category. The latter category takes network dynamics into account, and it includes attributes $a_5,a_{6-dynamic},a_{7-dynamic}$. We apply the best-first search based on the above definitions of weights, and investigate in Section \ref{5.2:pd} which of these weight types lead to a better extraction of important pathways.

\subsection{Pathway Score}
\label{4.3:ps}
With the algorithm explained in Section \ref{4.1:bfs} and weights definition in Section \ref{4.2:na}, we are now able to extract intervention pathways via graph search. Assuming that there is a regulatory pathway between a source node $x_s$ and a target node $x_t$, we assign a \textit{pathway score} to path:$\{x_s,x_{p_1},...,x_{p_m},x_t\}$ such that
\begin{equation} \label{eqn10}
s(path)=exp(-[w_{sp_1}+\sum_{i=1}^{m-1} w_{p_ip_{i+1}}+w_{p_mt}])
\end{equation}

The pathway score gives the influence of source node on the target node following that pathway. The optimal path (\textit{i.e.}, with minimum cost) has the highest path score. It is also clear that, with weight type $a_6(x_i,x_j)$ (\textit{i.e.},$w_{ij}=-log(a_6(x_i,x_j))$), computing $s(path)$ will return the element influence product. If we assume the independence between regulatory edges on a pathway, this product will indicate the propagated probability effect along the pathway, thus providing a measure of control from source node to target node.

\subsection{Pathway Extraction Evaluation}
\label{4.4:pee}
The underlying idea of our pathway extraction validation method is that the biological properties of the original model will not be preserved when an important pathway in the model (\textit{i.e.}, all of the interactions within this pathway) has been removed, while the removal of a pathway with low influence should not change the properties significantly. We use the biological property mismatch between the partial model after removal and the original baseline model, to verify our pathway extraction approach.

To conduct validation, we follow the algorithm described in Section \ref{4.1:bfs} and the weight definitions in Section \ref{4.2:na}. For each source-target pair of interest, and for each weight type, we define the pathway extraction threshold, that is, the number of top paths that we would like to extract and explore, assuming paths are ranked according to their impact. During the extraction, in case we encounter cycles, we allow at most one repeat for some pre-defined (\textit{e.g.}, according to prior knowledge) list of elements, and prohibit any other repeated element occurrences on the paths.
After obtaining top paths for each weight definition, we use the following procedure to validate the results of pathway extraction. First, we define a set of properties that are true for the modeled system. For example, a property can state a gene expression level at a particular time during observation interval, or relative changes in proteins in time. Next, we remove all edges (regulations) in the obtained top pathways from the original model to form a partial model. Finally, we use the statistical model checking approach (\citealp{13:misk16}, \citealp{14:wang16}), combined with DiSH simulations, to test all partial models on the defined set of properties, which returns the probabilities for each model satisfying each property.

In particular, when we remove the regulation of element $x_j$ by element $x_i$, we modify $x_j$'s update function such that $x_i$ is removed while keeping other elements and their original effects in the update function unchanged, which is, in some cases, achieved by setting $x_i$ to 0 or 1 permanently in the update function. In \textit{Example 1}, knocking down the regulation of element $x_1$ in $x_5=f_5=x_1 x_2+x_1^{'}x_3+x_2x_3^{'}x_4$ returns $x_5=f_5=x_2+x_3+x_2x_3^{'}x_4$. Note that this modification does not indicate the deletion of $x_1$ from our research scope, nor does it change the effect of $x_1$ on other elements than $x_5$.

\section{Results}
In Section \ref{5.1:cs}, we provide a brief background on T cell differentiation and the baseline model. In Section \ref{5.2:pd}, we confirm, using model checking and the T cell differentiation case study system, that our sensitivity analysis framework is successful in intervention pathway discovery. Finally, we study various features of the T cell differentiation circuitry using our proposed sensitivity analysis methods.

\subsection{Case Study: T-cell Differentiation}
\label{5.1:cs}
T cells, one of two primary types of lymphocytes, play a central role in cell-mediated immunity (\citealp{26:gutc07}). There are several subsets of T cells, and each one has a distinct function in the T-cell mediated immunity. In this work, we investigate T cell differentiation as our case study. Specifically, we use the model of the circuitry that controls differentiation of naive T cells into two types, regulatory cells (Treg), which suppress immune response and reduce the damage caused by autoimmune response, and helper cells (Th), which help promote immune response. Previous research (\citealp{26:gutc07}) has shown that these two types of T-cells are distinguished by expressions of several key elements. For example, in the Treg type, the transcription factor forkhead box P3 (Foxp3) is expressed, and Interleukin-2 (IL-2) is inhibited, while in the Th type, Foxp3 is inhibited and IL-2 is activated.

The model in (\citealp{15:misk13}) has overall 42 elements, including receptors TCR, CD28, TGF-$\beta$R, and IL-2R, genes for IL-2R$\alpha$ (\textit{i.e.}, CD25), IL-2, and Foxp3, and cytoplasmic members of PI3K/AKT/mTOR, MAPK, NF$\kappa$B, NFAT pathways. The authors in (\citealp{15:misk13}) used logical modeling approach to study naive T cell differentiation. In that work, several model elements are implemented as three-level discrete variables with values \{0,1,2\}, to denote absence, low activity, and high activity of the element, respectively. To accommodate three discrete levels within a logical model, two Boolean variables are associated with these elements with three levels. For example, the T-cell receptor (TCR) is modeled with two variables, TCR\_LOW and TCR\_HIGH, such that: TCR=0: (TCR\_LOW=0, TCR\_HIGH=0); TCR=1: (TCR\_LOW=1, TCR\_HIGH=0); TCR=2: (TCR\_LOW=0, TCR\_HIGH=1). The authors in (\citealp{15:misk13}) also use extra, dummy variables, to incorporate delays on several pathways, and therefore, the model in (\citealp{15:misk13}) has overall 55 logic variables. Here, we will focus on exploring actual element sensitivities, and will thus use element names instead of variable names. Further details of the variable implementation can be found in (\citealp{15:misk13}).

Besides the list of elements, and their update functions, as described in Section \ref{3.4:da}, we also need to define the scenarios for conducting dynamic sensitivity analysis on the T cell model. Similar to the studies in (\citealp{15:misk13}), we explore three dynamic scenarios: Scenario 1: high antigen dose (TCR=2), Scenario 2: low antigen dose scenario (TCR=1), and Scenario 3: a toggle scenario (TCR=2 initially, and changed to TCR=0 at a defined time step).

\subsection{Pathway Discovery}
\label{5.2:pd}
To demonstrate the use of our pathway discovery method, we start with the model from (\citealp{15:misk13}) and select two source-target pairs, (TCR, Foxp3) and (CD28, IL-2). Following the procedure described in Section \ref{4.4:pee}, for each of the nine network attributes defined in Section \ref{4.2:na}, and for each source target pair, we obtain three paths with minimum costs, computed according to Equation (\ref{eqn9}). We then create new models by removing these paths from the original model, and test their performance using model checking. While, in general, more complicated properties can be used, for our analysis of the T cell model, we test whether steady-state values of IL-2, Foxp3, AKT, and PTEN are reached. Specifically, we test whether elements (IL-2, Foxp3, AKT, PTEN) reach values (1,0,1,0) in Scenario 1, and (0,1,0,1) in Scenarios 2 and 3. We call the probability of satisfying a property, obtained from model checking, \textit{property match probability}.

\begin{figure}[thpb]
\centerline{\includegraphics[width=8.2cm]{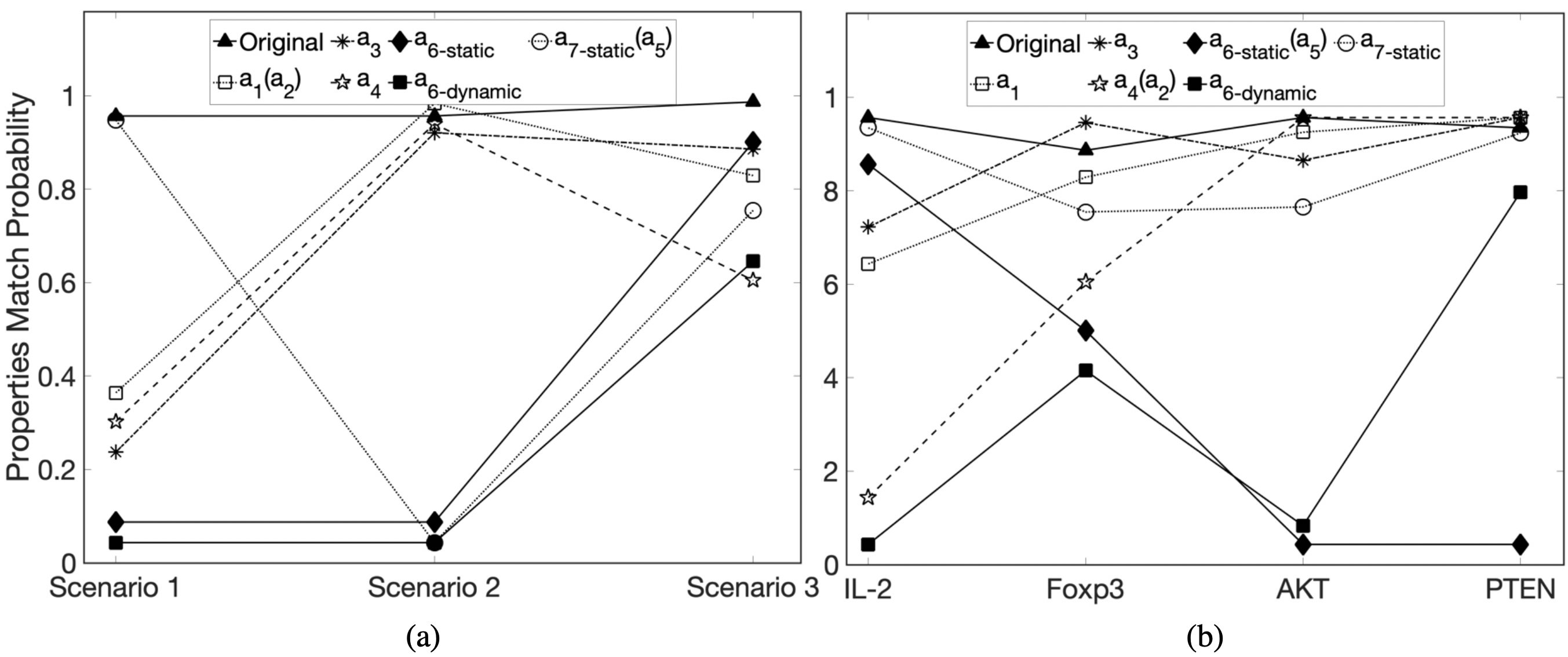}}
\caption{Probabilities of model property match for the original model and several partial models: (a) under three different scenarios; (b) under Scenario 2.}
\label{fig:06}
\end{figure}

For each model, we combine the property match probabilities into \textit{average match probability} by computing an average of these four probabilities. In Fig. \ref{fig:06}(a), we show the results for average match probability for several model versions under the three scenarios defined in Section \ref{5.1:cs}. The models that we compare are the original model (\citealp{15:misk13}) and partial models after removing pathways based on using different network attributes. As can be seen from Fig. \ref{fig:06}(a), in the original model (solid line with triangle marker), properties are highly satisfied (\textit{i.e.} average value of these four probabilities is close to 1). Note that we illustrate the results for the $a_1$-based and $a_2$-based removal using a single line in Fig. \ref{fig:06}(a), since their results are the same; similarly, we use a single line for $a_{7-static}$-based and $a_5$-based removal. It can be observed that all removals affect the property match probability, and moreover, while removing pathways based on in-degree or out-degree distribution ($a_1$ and $a_2$), minimum length ($a_3$), and loop count ($a_4$) affects the properties only to some extent under Scenarios 1 and 3, removing pathways based on element influence ($a_{6-static}$, $a_{6-dynamic}$) significantly breaks down the model (match probability approaches 0) under Scenarios 1 and 2.

\begin{figure}[thpb]
\centerline{\includegraphics[width=7.7cm]{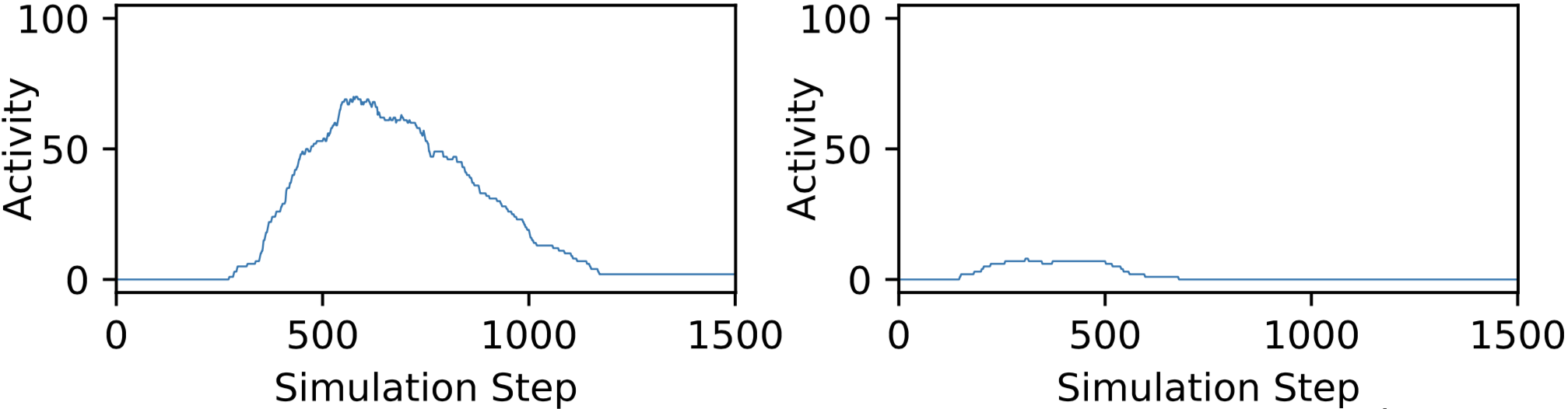}}
\caption{The activity of element IL-2: original model (left), and partial model after pathway removal based on $a_{6-dynamic}$ (right).}
\label{fig:07}
\end{figure}

In addition, the dynamic sensitivity analysis of the $a_{6-dynamic}$-based pathway extraction is conducted with respect to the distribution estimated under Scenario 2. Therefore, we explored further this particular scenario, and illustrated in Fig. \ref{fig:06}(b) these four detailed probabilities under Scenario 2. It is clear from the figure that removing pathways based on the dynamic influence value $a_{6-dynamic}$ calculated under Scenario 2, has an even more destructive effect on the model behavior under that scenario. By removing the pathways based on $a_{6-dynamic}$, three of the four key elements have a significant mismatch with their corresponding desired properties (solid line with triangle marker in the figure), thus revealing the critical role of the removed pathways. To further show the effect, we plot in Fig. \ref{fig:07} the change in time (trajectory averaged across 50 runs, each run with 1500 steps) of element IL-2 in the original model (left) and in the partial model (right) after we removed pathways using the $a_{6-dynamic}$ attribute.

\subsection{Element-level Analysis: Element Sensitivity}
\label{5.3:es}
By applying model checking and testing important system properties, we have shown in Section \ref{5.2:pd} that element influence and element sensitivity are good indicators of pathway importance. This is due to the fact that, the calculations of element influence and sensitivity in Equations (4, 5, 8) take both network structure and network dynamics into account. In other words, element influence and sensitivity can be used to define weights and incorporate context-dependent information when determining important pathways, thus enabling extension from local to global analysis.

\begin{figure}[thpb]
\centerline{\includegraphics[width=8cm]{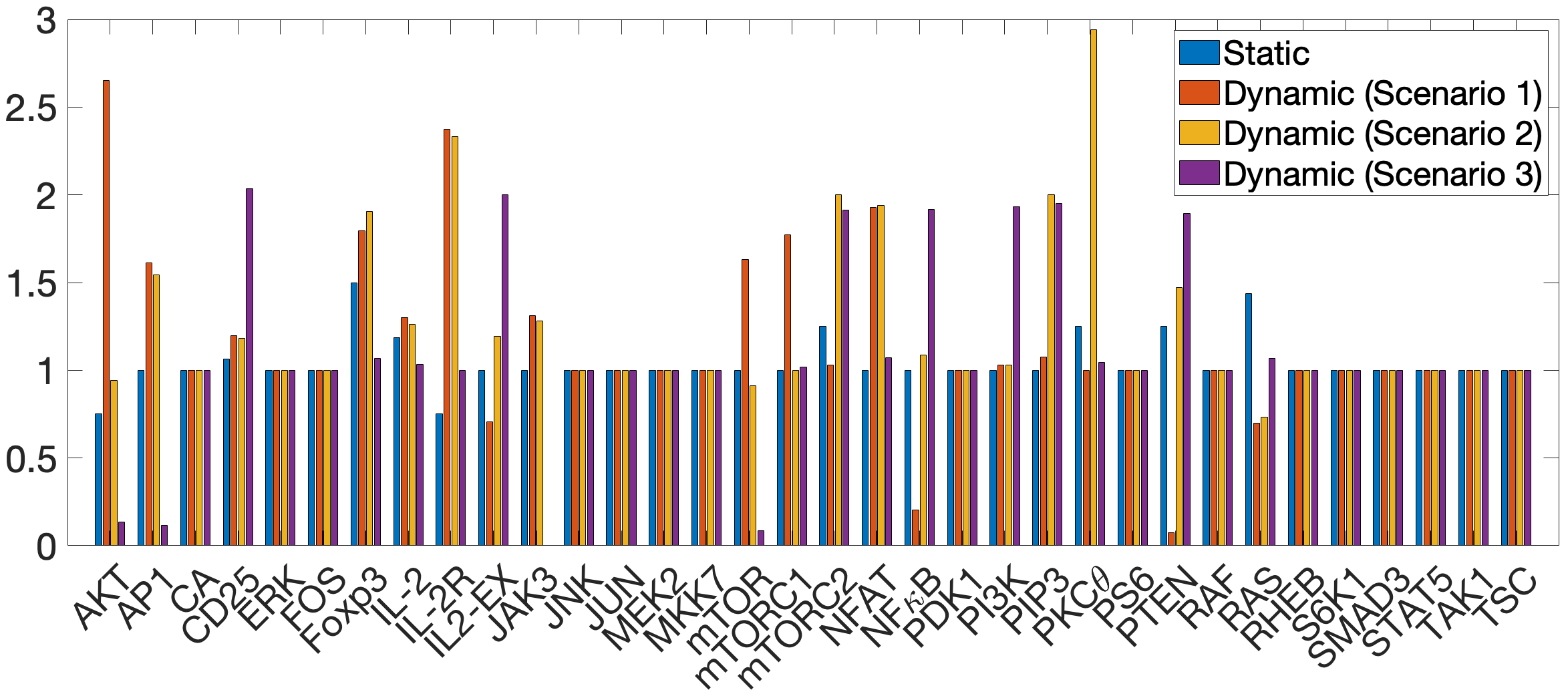}}
\caption{The sensitivity distribution of all elements under four cases of analysis.}
\label{fig:08}
\end{figure}

For the T cell model from (\citealp{15:misk13}), the interaction map $G(V,E)$ is fixed, while the weighted directed graph $G(V,E,W)$ varies with respect to state distribution, which is determined by simulation scenario. Under different scenarios, elements have different sensitivities. Fig. \ref{fig:08} shows the sensitivity of model elements using static analysis defined by Equation (\ref{eqn5}) in Section \ref{3.3:sa} (blue bars), as well as dynamic analysis defined by Equation (\ref{eqn8}) in Section \ref{3.4:da} under the three different scenarios (red, orange and purple bars for Scenario 1,2, and 3, respectively). Note that in Fig. \ref{fig:08} we omit results for elements that do not have update functions (CD28, TCR, TGF$\beta$, TGF$\beta$R, CD122, CD132), and therefore, cannot be sensitive to changes in any other model elements.

As can be seen from Fig. \ref{fig:08}, static analysis results in less variance, ranging from 0 to 1.5. Most elements have sensitivities less than 1, indicating that this network is likely to follow a stable and ordered structured behavior (\citealp{1:shmu04}). However, dynamic analysis (in all three scenarios) shows greater variance, ranging from 0 to 2.94. Some elements under dynamic analysis have sensitivities much greater than 1 (\textit{e.g.} element PKC$\theta$ in low-dose scenario). These results indicate that PKC$\theta$ is almost 100\% sensitive to any one-bit change of its regulators (\textit{i.e.}, TCR, CD28, mTORC2) and it will always propagate the change to its immediate downstream connections (\textit{i.e.}, NF$\kappa$B, TAK1). This kind of local ``instability'' can suggest potential intervention points on the pathways leading to phenotype markers Foxp3 and IL-2. Other elements which also show similar behavior include AKT, IL-2R in high-dose scenario, IL-2R, mTORC2, PIP3 in Scenario 2 (low-dose), and CD25, IL-2-exogenous (IL2-EX) in Scenario 3 (toggle).

Additionally, it can be observed from the results in Fig. \ref{fig:08} that the top elements behaving differently across the three scenarios are AKT and PKC$\theta$. For example, AKT shows robustness against changes in its regulators PDK1 and mTORC2 in Scenario 3 (toggle), while exhibits sensitivity to these regulators in Scenario 1 (high-dose). In other words, high-dose stimulation makes element AKT inclined towards the states that are highly sensitive and flexible against environment changes. This feature is quite useful when designing interventions to efficiently induce certain system response given particular initial state, or for identifying elements with critical role in the context-dependent analysis. Finally, this also emphasizes the importance of the context-dependent analysis itself.

\subsection{Interaction-level Analysis: Element Influence}
\label{5.4:ei}
\begin{figure}[thpb]
\centerline{\includegraphics[width=8cm]{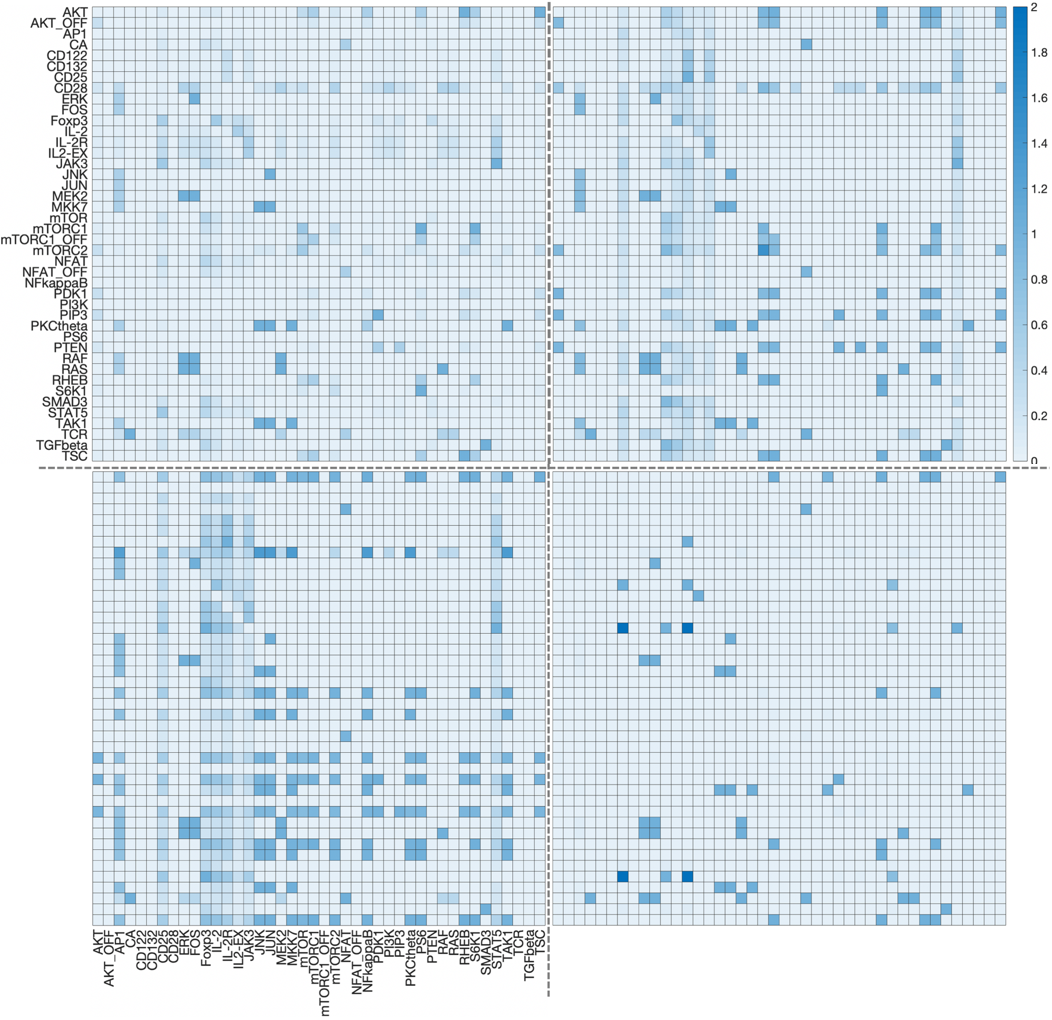}}
\caption{The element-to-element influence matrix under four cases: static analysis (top left), and the three dynamic analysis scenarios (top right and bottom left and right)}
\label{fig:09}
\end{figure}

When designing models of systems on the basis of incomplete information, and moreover, when defining intervention strategies for such systems, it is beneficial to also conduct analysis at an element-to-element (interaction) level. This type of analysis can summarize the details of indirect regulations into a single measure, \textit{element-to-element influence}. Using this measure, we can study influence of an arbitrary source element, $x_s$, on an arbitrary target element, $x_t$. First, we find all paths from $x_s$ to $x_t$, and then we summarize the score of all the paths as element-to-element influence. As defined in Equation (\ref{eqn10}), a single pathway score only implies the influence following that pathway. If we add the scores of all possible paths from $x_s$ to $x_t$, this summation shows the overall influence that the source element $x_s$ has on the target element $x_t$. In Fig. \ref{fig:09}, we show the element-to-element influence matrices under four cases. Rows in the heatmaps correspond to source elements and columns correspond to target elements.

In general, element-to-element influences in three dynamic scenarios are greater than element-to-element influences under static analysis. There are two main reasons for this. First, under static analysis, the long-run element-to-element influences are quite sensitive to the length of pathways since the direct element influence $\alpha$ under static analysis is relatively small. Second, dynamic scenarios are obtained from real biological observations and thus show stronger homogeneity.

Additionally, it is worthwhile to note that, compared to the static analysis, the influence matrices of the three dynamic scenarios shows significant imbalance (\textit{i.e.}, with wide variance range and high deviation), especially the matrix of toggle case. It has been shown before that systems with unbalanced influence distributions are stable and robust (\citealp{9:anon07}). Thus, the results in Fig. \ref{fig:09} suggest that we may toggle node TCR in the initial state, or toggle nodes like PKC$\theta$ and PI3K during the transient process in order to drive the system to certain states or to maintain system stability.

Intuitively, a high element-to-element influence (darker blue blocks) indicates strong interaction. It is easy to observe examples of such strong interactions from Fig. \ref{fig:09}, such as CD28 to {JUN, MKK7} in low-dose scenario, and {JAK3, STAT5} to {CD25, IL-2R} in toggle scenario.

\subsection{Static vs. Dynamic Analysis}
\label{5.5:sd}
Scenario specifics (\textit{i.e.}, context) are often critical for guiding intervention decisions. For example, detailed information about the effects of a scenario on the model helps evaluate the trade-off between accuracy and effort when designing context-specific vs. generalized interventions.

We have observed that some pathways that appear very influential in the static analysis are actually inactive in dynamic scenarios. These observations suggest that adding interventions following only static pathway and element function analysis may not be effective.

To demonstrate the significance of change in the number of active pathways when switching between static and dynamic analysis, quite common for many source/target combinations, Table \ref{Tab:01} gives the comparison in number of active pathways under static analysis and dynamic analysis with high-dose scenario and low-dose scenario.
\begin{table}[!t]
\caption{Comparison in number of active pathways in different cases}
\label{Tab:01}
\centerline{\includegraphics[width=7.7cm]{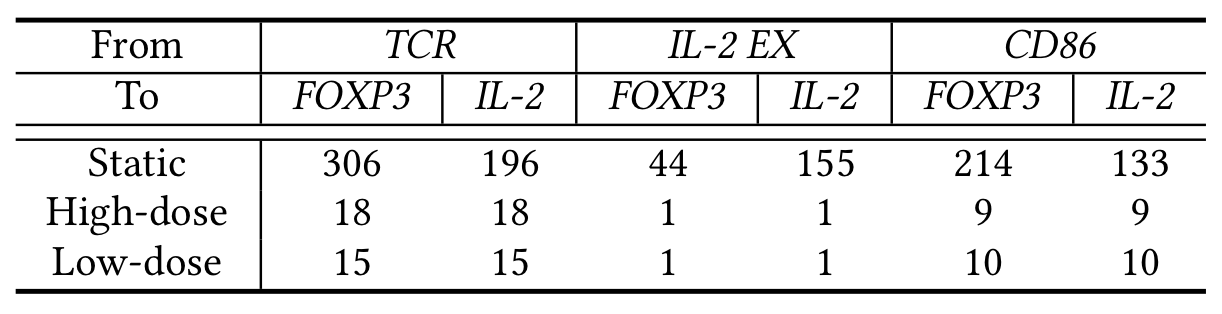}}
\end{table}

As can be seen from Table \ref{Tab:01}, accounting for the real scenarios that can occur in the cell (scenarios within our dynamic analysis) leads to reduction in the number of active pathways. More interestingly, we find that there is only one active pathway from IL-2 EX to IL-2 in high-dose/low-dose scenarios, which is IL-2 EX, JAK3, STAT5, Foxp3, IL-2. Also, in dynamic case, the number of active pathways of regulation for Foxp3 and IL-2 does not change with the change of a scenario, and this holds for all source nodes.

\section{Conclusion}
In this work, we model and investigate intra-cellular networks via discrete modeling approach, and we propose a framework to study sensitivity in these models.  Previous sensitivity analysis assumes uniform state distribution, which is not always true in biology. We perform both static and dynamic sensitivity analysis, the former assuming uniform state distribution, and the latter using a distribution estimated from stochastic simulation trajectories for a given scenario. Within our sensitivity analysis framework, we first compute element-to-element influences, then we extend the element update functions to include weights according to these computed influences. Adding weights to these interaction rules helps to identify key elements in the model, as well as dominant signaling pathways that determine the behavior of the overall model. In the end, we apply our sensitivity analysis framework on the intervention pathway extraction and evaluation in the intra-cellular network that controls T cells differentiation. Our results emphasize the importance of incorporating context in sensitivity analysis and in the selection of intervention pathways.


\section*{Funding}
This work is supported in part by DARPA award W911NF-17-1-0135 and W911NF-18-1-0017.

\end{document}